\documentclass[conference]{IEEEtran}
\usepackage{amsmath}
\usepackage{cite}
\usepackage{amssymb}
\usepackage{amsfonts}
\usepackage{algorithm}
\usepackage{algpseudocode}
\usepackage{dsfont}
\usepackage{graphicx}
\usepackage{epsfig}
\usepackage{subfigure}
\usepackage{psfrag}
\usepackage{xcolor}
\usepackage{url}
\usepackage[colorlinks,linkcolor=black,urlcolor=black,anchorcolor=black,citecolor=black,hyperfootnotes=true]{hyperref}
\usepackage{tikz}
\usepackage{stfloats}

\usepackage{hyperref}
\hypersetup{draft}

\usepackage{enumitem}
\usepackage{setspace}

\newcommand{\mv}[1]{\mbox{\boldmath{$ #1 $}}}

\title{Rethinking Grant-Free Protocol in mMTC}
\author{Minhao Zhu$^{\S}$, Yifei Sun$^{\S}$, Lizhao You$^{\star}$, Zhaorui Wang$^{\S}$, Ya-Feng Liu$^{\ddag}$, and Shuguang Cui$^{\S}$\\$^{\S}$FNii and SSE, The Chinese University of Hong Kong (Shenzhen), Shenzhen, China\\
$^{\star}$School of Informatics, Xiamen University\\
$^{\ddag}$LSEC, ICMSEC, AMSS, Chinese Academy of Sciences, Beijing, China\\Email: wangzhaorui@cuhk.edu.cn}
\begin{document}
\maketitle

\begin{abstract}
This paper revisits the identity detection problem under the current grant-free protocol in massive machine-type communications (mMTC)  by asking the following question: for stable identity detection performance, is it enough to permit active devices to transmit preambles without any handshaking with the base station (BS)? Specifically, in the current grant-free protocol, the BS blindly allocates a fixed length of preamble to devices for identity detection as it lacks the prior information on the number of active devices $K$. However, in practice,  $K$ varies dynamically over time, resulting in degraded identity detection performance especially when $K$ is large. Consequently, the current grant-free protocol fails to ensure stable identity detection performance. To address this issue, we propose a two-stage communication protocol which consists of estimation of $K$ in Phase I and detection of identities of active devices in Phase II. The preamble length for identity detection in Phase II is dynamically allocated based on the estimated $K$ in Phase I through a table lookup manner such that the identity detection performance could always be  better than a predefined threshold. In addition, we design an algorithm for estimating $K$ in Phase I, and  exploit the estimated $K$ to reduce the computational complexity of the identity detector in Phase II. Numerical results demonstrate the effectiveness of the proposed two-stage communication protocol and algorithms. 
\end{abstract}


\section{Introduction}
Massive machine-type communications (mMTC) is envisioned as one of the candidates in 6G. mMTC provides efficient random access communications for a large number of devices, out of which only a small number of them are active. To reduce communication latency, a grant-free communication protocol is adopted where devices can transmit signals without permission from the base station (BS)\cite{bockelmann2016massive,senel2018grant,LIU2023}. Each device under the grant-free protocol is pre-assigned with a unique and nonorthogonal preamble that is served as identity of the device. A significant challenge posed by the grant-free protocol is to detect the identities of active devices through the transmission of  preambles\cite{bockelmann2016massive,LIU2023}. The identities of active devices are useful in subsequent stages, e.g., resource allocation and channel estimation\cite{senel2018grant,wang2023device}. 

This paper revisits the identity detection problem under the current grant-free protocol in mMTC by asking the following question: for stable identity detection performance, is it enough to permit the active devices to transmit preambles without any handshaking with the BS? Specifically, in the current grant-free protocol, since the BS has no prior information on the number of active devices $K$, it just blindly allocates a fixed length of preamble $L_{\rm II}$ for identity detection, no matter how large the actual value $K$ is. However, in practice, $K$ varies dynamically over time.
Theoretical findings from \cite{Chenphase} suggest that maintaining a fixed $L_{\rm II}$ at the BS significantly compromises identity detection performance when $K$ is large. This scenario could lead to identity detection performance falling below a predefined threshold, which is undesirable in practical communication systems.

\begin{figure}[t]
 	\centering
 	\includegraphics[width=5cm]{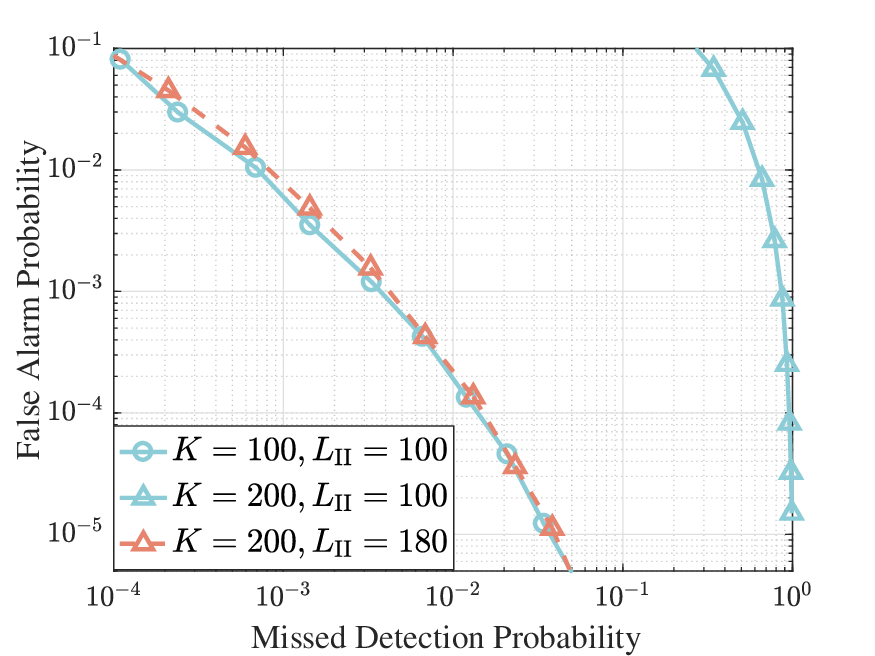}
 	\caption{Identity detection performance in the scenario where the total number of devices $N=1000$  and the number of BS antennas $M=15$.} \label{diffK1}
\end{figure}

We take Fig. \ref{diffK1} as an example for further illustration, where  a covariance-based approach in \cite{haghighatshoar2018improved} is applied to detect the identities of active devices. Fig. \ref{diffK1} shows that for fixed $L_{\rm II}=100$, the identity detection performance decreases significantly when the number of active devices $K$ increases from 100 to 200. If we improve the preamble length to $L_{\rm II}=180$, the identity detection performance when $K=200$ approaches to that when $K=100$. This phenomenon underscores the necessity of dynamically allocating the preamble length $L_{\rm II}$ according to the dynamic value of $K$, rather than blindly assigning a fixed length.

Building upon this observation, we argue that the current grant-free protocol is inadequate for ensuring stable identity detection performance. In particular,  the number of active devices $K$ is required for the allocation of preamble length $L_{\rm II}$. To solve this problem, we first propose a two-stage communication protocol, which consists of estimation of $K$ in Phase I and detection of identities of active devices in Phase II. The preamble length $L_{\rm II}$ for identity detection in Phase II is dynamically allocated based on the estimated $K$ in Phase I through a table lookup manner, such that the identity detection performance could  always be  better than a predefined threshold. Second, in Phase I of the protocol, we propose an efficient algorithm to estimate $K$ at a negligible cost of preamble symbols. Third, in addition to the allocation of preamble length $L_{\rm II}$, we found that the estimated $K$ can also be exploited to reduce the computational complexity of the identity detector in Phase II. We propose an efficient algorithm to achieve this goal. Numerical results demonstrate the effectiveness of the proposed two-stage communication protocol and algorithms. 

Previous works, including \cite{senel2018grant,Chenphase,haghighatshoar2018improved,sun2019exploiting,jiang2018joint,ke2020compressive,mei2021compressive,wang2022covariance,ganesan2020algorithm,lin2022sparsity,wang2023cov}, have primarily focused on the design of identity detector under the grant-free protocol but have overlooked the importance of ensuring stable identity detection performance, which is crucial in practical systems.  In contrast to these works, we study a two-stage communication protocol that estimates $K$ before the identity detector in order to achieve stable identity detection performance. In addition,  \cite{wang2023cov} has proposed an algorithm to reduce the computational complexity of the identity detector. Compared with \cite{wang2023cov}, our proposed algorithm is much more efficient to reduce the computational complexity by leveraging the knowledge of the estimated $K$.

\section{Two-Stage Communication Protocol}\label{s2}
We consider a narrowband uplink system which consists of a BS equipped with $M$ antennas and $N$ single-antenna devices. At each time slot, only $K \ll N$ devices are active. Denote $\gamma_{n}$ an activity indicator of device $n$, i.e., 
\begin{equation}
	\gamma_n=\left\{\begin{array}{l}
		1, \text { if device } n \text { is active, }\\
		0, \text { otherwise, }
	\end{array}\right.
	\begin{aligned}
		\quad n=1,2, \ldots ,N.
	\end{aligned}
\end{equation}
In the traditional grant-free protocol, each device is pre-assigned with a unique and nonorthogonal preamble with a fixed length of $L_{\rm II}$  to detect the identities of active devices through the estimation of $\gamma_n$, $n=1, 2, \dots, N$. 

\subsection{Problem of Existing Grant-Free Protocol}
In this paper, we take the covariance-based approach as an example to show the problem of identity detection under the current grant-free protocol. Specifically, \cite{Chenphase} studied under what conditions the identities of all $K$ devices can be correctly detected by the BS. It has been proved in \cite{Chenphase} that when $M$ is sufficient large,  the maximum number of active devices that can be correctly detected scales as
\begin{align}
  K\sim\mathcal{O}\left(L_{\rm II}^2\right), \label{eq:od}
\end{align}
where $\mathcal{O}$ denotes the order of approximation. Eq. \eqref{eq:od} clearly shows that the maximum number of active devices $K$ that can be correctly detected highly depends on the preamble length $L_{\rm II}$, although the exact expression of  $K$ with respect to $L_{\rm II}$ is not clear far from now. On the one hand, under the current grant-free protocol, since there is no handshaking between the devices and the BS before the signal transmission, the BS has no prior information of $K$ at each time slot. In this case, the BS can only blindly allocate a fixed length of preamble $L_{\rm II}$ for identity detection. On the other hand, the number of active devices $K$ varies at different time slots. In the heavy-load case where $K$ is far larger than $L_{\rm II}^2$, the identity detection performance would be much worse. 

In practical scenarios, to ensure system-level performance, a practical system typically defines a threshold that delineates the minimum acceptable performance for identity detection across the entire system. Through the aforementioned analysis, it becomes evident that the current grant-free protocol may yield identity detection performance inferior to the predefined threshold, thereby degrading the overall system performance.

\subsection{Proposed New Communication Protocol}
To solve the above problem, the BS should have the prior information on the number of active devices $K$. In this case, the BS can allocate the preamble length $L_{\rm II}$ for identity detection according to the estimated $K$. However, the existing literature has not established an explicit relationship between $K$ and $L_{\rm II}$ for a specific identity detection error rate when $M$ is finite.

To solve the problem of allocating $L_{\rm II}$ according to the estimated $K$, we propose to create a table that specifies this relationship through simulations. Once we have the knowledge of estimated $K$, we can determine the preamble length $L_{\rm II}$ through a table lookup manner. The table lookup approach has been widely applied in wireless systems when the explicit relationship among the parameters is hard to characterize, e.g., parameter selection of  channel coding and modulation in 5G\cite{3P}.  

The overall proposed two-stage communication protocol is summarized as follow: 
\begin{enumerate}
	\item \textbf{Phase I: Estimation of the number of active devices $K$}. Device $n$ is allocated $L_{\rm I}$ preamble symbols to estimate the number of active devices $K$ at the BS, $n=1,2,\dots, N$. Then, based on the estimated $K$, the BS allocates $L_{\rm II}$ preamble symbols through a table lookup manner, in order to detect the identities of active devices.
	\item \textbf{Phase II: Detection of the identities of active devices}. The BS detects the identities of active devices based on the received $L_{\rm II}$ preamble signals from all active devices.
\end{enumerate}
The overall preamble length is
\begin{align}
	L=L_{\rm I}+L_{\rm II}.\label{eq:ov}
\end{align}

There are several technical problems to be addressed in the proposed communication protocol. First, how to efficiently estimate $K$? That is, if we spend too many preamble symbols on the estimation of $K$ in Phase I, the time for identity detection in Phase II would be quite limited. Second, the proposed estimation method for $K$ in Phase I should not be sensitive to $K$ itself; otherwise, we should also allocate $L_{\rm I}$ dynamically according to the dynamic value of $K$, which further complicates the design of the communication protocol. Third, although the primary goal of estimating $K$ is to determine the length of preamble $L_{\rm II}$ for identity detection, it's important to explore how to further exploit the estimated $K$  to improve the performance of the identity detector in Phase II? We will address the above problems in the rest of this paper.

\section{Estimation of Number of Active Devices} \label{sec:P1}
In this section, we show how to estimate the number of active devices $K$. In addition, through numerical results, we show that the parameter $K$ can be estimated through only a few preamble symbols, and the proposed estimator is not sensitive with $K$, i.e., the estimator keeps roughly the same performance under different $K$.

\subsection{Problem Formulation and Estimator Design}
In Phase I, in order to estimate $K$, all devices apply a same preamble $\mv{s}\in\mathbb{C}^{L_{\rm I}\times1}$. In addition,  we assume block-fading channels, i.e., the channels among the devices and BS remain roughly constant within each coherence block, but may vary among different coherence blocks. The channel from device $n$ to BS is denoted by $\sqrt{\beta_n}\mv{h}_n\in \mathbb{C}^{M\times1}$, where $\beta_n$ denotes the large-scale fading component, and $\mv{h}_n\sim\mathcal{CN}(0,\mv{I})$ denotes the independent and identically distributed (i.i.d) Rayleigh fading coefficient. Assume active devices transmit their preambles at a predefined time slot such that the received signals are perfectly synchronized at the BS. In this case, the received signal at Phase I becomes 
\begin{align}
	\small{\bar{\mv{Y}}=\sum_{n=1}^{N}\gamma_n\sqrt{p_n}\mv{s}\sqrt{\beta_n}\mv{h}^T_n+\bar{\mv{Z}}\overset{(a)}{=}\sqrt{\beta}\mv{s}\sum_{n=1}^{N}\gamma_n\mv{h}^T_n+\bar{\mv{Z}}},\label{eq:rec25}
\end{align}
where $p_n$ is the transmit power of device $n$, and $\bar{\mv{Z}}$ is the additive white Gaussian noise,  (a) is because we do power-control  on each device, i.e., 
\begin{align}
	p_n\beta_n=\beta,~n=1, 2,\dots, N.\label{eq:beta2}
\end{align}
The received signal $\bar{\mv{Y}}$ is then normalized by $\sqrt{\beta}$, and is expressed as
\begin{align}
	\small{\mv{Y}=\bar{\mv{Y}}/\sqrt{\beta}=\left[\mv{y}_{1}, \mv{y}_{2}, \dots,\mv{y}_{M}\right]=\mv{s}_{}\sum_{n=1}^{N}\gamma_n\mv{h}^T_n+\mv{Z}},\label{eq:rec2}
\end{align}
where $\mv{y}_{m}$ is the received signal at the $m$-th antenna,  $\mv{Z}=\bar{\mv{Z}}/\sqrt{\beta}=\left[\mv{z}_{1}, \mv{z}_{2}, \ldots, \mv{z}_{M}\right]\in\mathbb{C}^{L_{\rm I}\times{M}}$ is the normalized additive white Gaussian noise with $\mv{z}_{m} \sim \mathcal{C} \mathcal{N}\left(0, \sigma^2 \mv{I}\right)$, $m=1, 2, \dots, M$, and $\sigma^2$ is the normalized noise power. In \eqref{eq:rec2}, since the channels follow the i.i.d. Rayleigh distribution, we have
\begin{align}
	\sum_{n=1}^{N}\gamma_n\mv{h}^T_n\overset{d}{=}\sqrt{K}\mv{h}^T_1\sim \mathcal{CN}(0,K\mv{I}),
\end{align}
where $\overset{d}{=}$ denotes equal in  distribution. In this case,  we have 
\begin{align}
	\mv{Y}\overset{d}{=}\sqrt{K}\mv{s}\mv{h}^T_1+\mv{Z}.\label{eq:rec3}
\end{align}
That is, the received signal $\mv{Y}$ is a function of $K$ in terms of its distribution. Denote $p(\mv{Y}
\mid K,\mv{s})$ the probability density function of $\mv{Y}$ given $K$ and $\mv{s}$, and it is expressed as
\begin{align}
	\small{p(\mv{Y}
	\!\!\mid \!\!K,\mv{s})\!\!=\!\!\prod_{m=1}^M p\left(\mv{y}_{m} \mid K,\mv{s}\right) \!\!=\!\! \frac{1}{|\mv{\Sigma}\pi|^M} \prod_{m=1}^M \exp \left(-\mv{y}_{m}^H \mv{\Sigma}^{-1} \mv{y}_{m}\right),}
\end{align}
where 
\begin{align}
	\mv{\Sigma}=\mathbb{E}\left(\mv{y}_{m} \mv{y}_{m}^H\right)  
	=K\mv{s} \mv{s}^H+\sigma^2 \mv{I}.
\end{align}
Note that maximizing $p(\mv{Y}
\mid K,\mv{s})$ is equivalent to minimizing $-\log\left(p(\mv{Y}
\mid K,\mv{s})\right)$. In this case, the estimation problem is thus formulated as
\begin{alignat}{3}
	&\min_{K} &\quad &\log|\mv{\Sigma}|+\rm{Tr}(\mv{\Sigma}^{-1}\hat{\mv{\Sigma}}) \label{eq:P4}\\
	&~~\text{s.t.}                          &      & N\ge K\ge 0,\nonumber
\end{alignat}
where 
\begin{align}
	\hat{\mv{\Sigma}}=\frac{1}{M}\sum_{m=1}^{M}\mv{y}_{m} \mv{y}_{m}^H.
\end{align}
By taking derivative on the objective function in \eqref{eq:P4} with respect to $K$, we can get the estimated $K$ as
\begin{equation}
	\hat{K}=\frac{\mv{s}^H \hat{\mv{\Sigma}}\mv{s}}{\left\|\mv{s}\right\|_2^4}-\frac{\sigma^2}{\left\|\mv{s}\right\|_2^2}.\label{K}
\end{equation}
\begin{figure}[t]
	\centering
	\subfigure[Estimation performance over $L_{\rm I}$ when $K=100$, where the error bar denotes the variance of the normalized estimation error.]
	{
		\label{K2}
		\includegraphics[width=0.55\columnwidth]{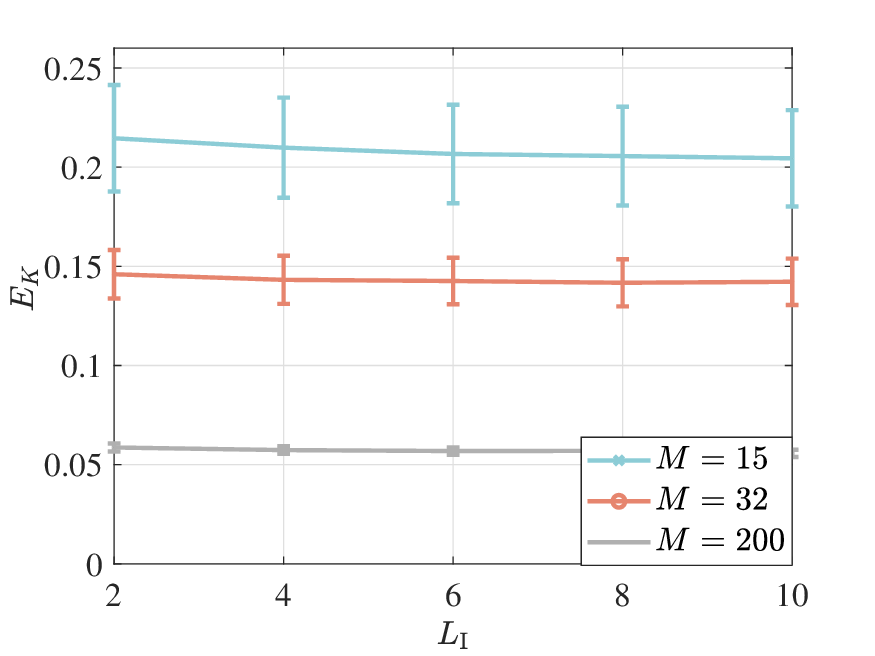}
	}
	\subfigure[Estimation performance when $L_{\rm I}=4$.]
	{
		\label{K1}
		\includegraphics[width=0.55\columnwidth]{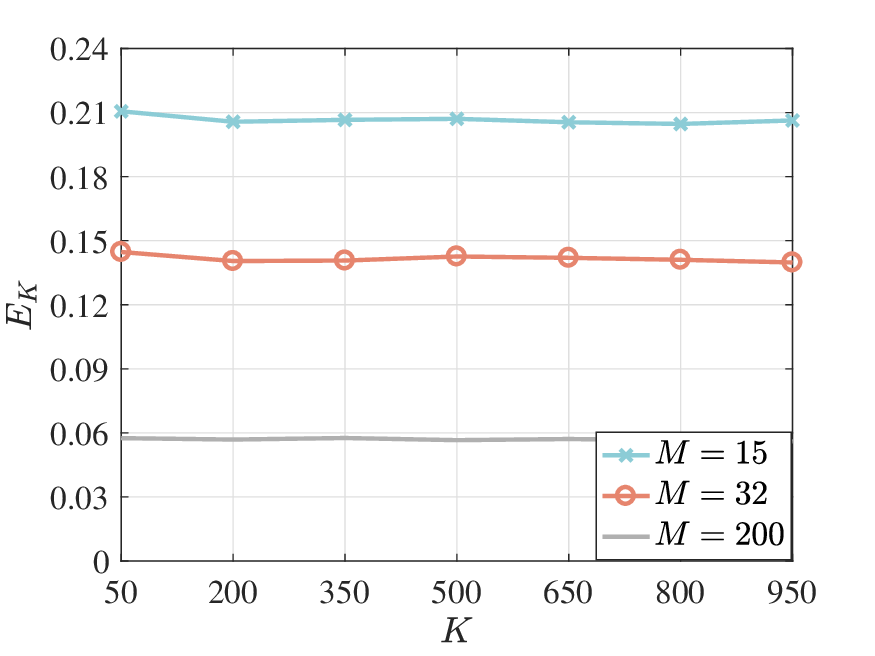}
	}
	\caption{Estimation performance over different $L_{\rm I}$ and $K$ when $N=1000$.}\label{KK}
\end{figure}

\subsection{Performance of Designed Estimator}\label{sec:D}
We present the performance of the proposed estimator when $N=1000$. Denote $E_K$ as the normalized estimation error, i.e.,
\begin{align}
	E_K=\mathbb{E}\left(\Big{|}\frac{K-\hat{K}}{K}\Big{|}\right),
\end{align}
where $\mathbb{E}(x)$ denotes the expectation of $x$. The noise power spectrum density is -169 dBm/Hz, the bandwidth is 10 MHz, the transmit power of each device is set to be 23 dBm, and  the large-scale fading component is modeled as  $128.1+37.6\log_{10}(d)$, where $d$ is the distance between the BS and devices in km\cite{3GPP}.  In the simulations, we set $d=1$ km. 

The numerical results are shown in Fig. \ref{KK}.  First, Fig. \ref{K2} shows that the estimation error decreases slowly with the preamble length $L_{\rm I}$. In this case, we can set a small value of $L_{\rm I}$ for the estimation of $K$, e.g., $L_{\rm I}=4$. Second, Fig. \ref{K1} shows that the estimation error keeps roughly the constant under different $K$, which suggests that our estimator is robust to $K$. In this case, we can set a same $L_{\rm I}$ to estimate $K$, regardless the value of $K$. Third, increasing the number of antennas $M$ at BS can reduce the estimation error significantly. 

We will have a further evaluation of the estimation error of $K$ on the two-stage protocol and the identity detector in Section \ref{Results}. 

\section{Detection of Identities of Active Devices}\label{sec:PhII}

In this section, we show how to reduce the computational complexity of the identity detector in Phase II by exploiting the information $\hat{K}$ obtained in Phase I.

\subsection{Problem Formulation}
In Phase II,  device $n$ is allocated a preamble $\mv{s}_{n}\in\mathbb{C}^{L_{\rm II}\times1}$ to report its activity to the BS, $n=1,2,\dots, N$. The channel  and noise models are the same as those in \eqref{eq:rec2}. The received signal at the BS is expressed as
\begin{align}
	\small{\mv{Y}=\!\!\left[\mv{y}_{1}, \mv{y}_{2}, \dots,\mv{y}_{M}\right]\!\!=\!\!\sum_{n=1}^{N}\gamma_n\mv{s}_{n}\mv{h}^T_n+\mv{Z}\!\overset{(b)}{=}\!\!\mv{S}\mv{\gamma}^{\frac{1}{2}}\mv{H}+\mv{Z},} \label{eq:rec5}
\end{align}
where $\mv{y}_{m}$ is the received signal at the $m$-th antenna. In \eqref{eq:rec5}, we reuse the notations $\mv{Y}$, $\mv{y}_{m}$, and $\mv{Z}$ in this subsection in order to ease the exposition; (b) in \eqref{eq:rec5} follows as we denote $\mv{S}=\left[\mv{s}_{1}, \mv{s}_{2}, \dots, \mv{s}_{N}\right] \in \mathbb{C}^{L_{\rm II}\times N}$  the preamble sequence matrix; $\mv{\gamma}=\operatorname{diag}\left(\gamma_1, \gamma_2, \ldots,\gamma_N\right) \in \mathbb{R}^{N \times N}$; and $\mv{H}=\left[\mv{h}_1, \mv{h}_2, \ldots, \mv{h}_N\right]^T \in \mathbb{C}^{N \times M}$. 
 
Denote $p(\mv{Y} \mid \mv{\gamma}, \mv{S})$ the PDF of $\mv{Y}$ given $\mv{\gamma}$ and $\mv{S}$. Then, we get
\begin{align}
	p(\mv{Y} \mid \mv{\gamma}, \mv{S})
	& = \frac{1}{|\mv{\Sigma} \pi|^M} \prod_{m=1}^M \exp \left(-\mv{y}_{m}^H \mv{\Sigma}^{-1} \mv{y}_{m}\right) \nonumber\\
	& = \frac{1}{|\mv{\Sigma} \pi|^M} \exp \left(-\operatorname{Tr}\left(\mv{\Sigma}^{-1} \mv{Y} \mv{Y}^H\right)\right),\label{pY}
\end{align}
where 
\begin{equation}
\mv{\Sigma}=\mv{S} \mv{\gamma} \mv{S}^H+\sigma^2 \mv{I}=\sum_{n=1}^N \gamma_n \mv{s}_{n} \mv{s}_{n}^H+\sigma^2 \mv{I}.\label{covmatrix}
\end{equation}
Note that maximizing $p(\mv{Y} \mid \mv{\gamma}, \mv{S})$ is equivalent to minimizing $-\log(p(\mv{Y} \mid \mv{\gamma}, \mv{S}))$. In this case, the detection problem is thus formulated as
\begin{alignat}{3}
	&\min_{\mv{\gamma}} &\quad &\log |\mv{\Sigma}|+\operatorname{Tr}\left(\mv{\Sigma}^{-1} \hat{\mv{\Sigma}}\right)  \label{eq:P6}\\
	&~~\text{s.t.}                          &      & \gamma_n\in[0,1], ~n=1, 2, \dots, N,\label{eq:cons}
\end{alignat}
where $\hat{\mv{\Sigma}}=\frac{1}{M} \mv{Y} \mv{Y}^H$ is the sample covariance matrix of the received signal. In the above, we reuse the notations $\mv{\Sigma}$ and $\hat{\mv{\Sigma}}$ in this subsection in order to ease the exposition. 

Our goal is to find a feasible point that satisfies
the first-order optimality condition of problem in \eqref{eq:P6}. To this end, denote the objective function in \eqref{eq:P6} by
\begin{align}
	f(\mv{\gamma})=\log |\mv{\Sigma}|+\operatorname{Tr}\left(\mv{\Sigma}^{-1} \hat{\mv{\Sigma}}\right). \label{eq:b}
\end{align}
Then, the gradient of $f(\mv{\gamma})$ with respect to $\gamma_n$ is given by
\begin{align}
	\left[\nabla f(\mv{\gamma})\right]_n=\mv{s}_{n}^H \mv{\Sigma}^{-1}\mv{s}_{n}-\mv{s}_{n}^H \mv{\Sigma}^{-1} \mv{\hat{\Sigma}}\mv{\Sigma}^{-1} \mv{s}_{n}. \label{grad equa2}
\end{align}
The first-order optimality condition of problem \eqref{eq:P6} is 
\begin{align}
	\small{\left[\nabla f(\mv{\gamma})\right]_n\left\{\begin{aligned}
		\geq 0,\  & ~~{\rm if}~\gamma_n=0, \\
		\leq 0,\  & ~~{\rm if}~\gamma_n=1, \\
		=0,\  & ~~{\rm if}~0<\gamma_n<1. 
	\end{aligned}\right. }\label{eq:grad}
\end{align}
To quantify the degree to which the coordinate $\gamma_n$ violates the optimality condition in \eqref{eq:grad}, in this paper, we define a nonnegative metric as follows:
\begin{equation}
	\small{V(\gamma_n)=\left\{\begin{aligned}
		&\Big|P_0^{+\infty}\big(-\left[\nabla f(\mv{\gamma})\right]_n\big)\Big|, ~~~~~~~{\rm if}~\gamma_n=0, \\
		&\Big|P_{-\infty}^1\big(1-\left[\nabla f(\mv{\gamma})\right]_n\big)-1\Big|,~{\rm if}~\gamma_n=1, \\
		&\Big|[\nabla f(\mv{\gamma})]_n\Big|,~~~~~~~~~~~~~~~~~~~~~{\rm if}~ 0<\gamma_n<1, 
	\end{aligned}\label{V2}\right.}
\end{equation}
where 
\begin{equation}
	\small{	P_a^b\left(x\right)=\left\{\begin{aligned}
		&a, ~~{\rm if}~x<a, \\
		&b,~~{\rm if}~x>b, \\
		&x,~~{\rm if}~ a\le x\le b. 
	\end{aligned}\label{V}\right.}
\end{equation}
Note that the first-order optimality is equivalent to
\begin{align}
	V(\gamma_n)=0, ~~n=1,2,\dots, N, 
\end{align}
where a larger $V(\gamma_n)$ indicates a larger violation of the first-order optimality.
 
\begin{algorithm}[t]
	\caption{CD Algorithm in \cite{haghighatshoar2018improved}}\label{alg:RCD}
	\begin{algorithmic}[1]
		\State Input: $\mv{\gamma}=\mv{0}$,  $\|\mv{V}\left(\mv{\gamma}\right)\|_{\infty}$, $\mv{\Sigma}=\sigma^2 \mv{I}$, $\varepsilon>0$.
		\While{$\|\mv{V}\left(\mv{\gamma}\right)\|_{\infty}>\varepsilon$}
		\For{$n=1,2,\ldots,N$}\label{l:l0}
		\State Update $\gamma_n$ according to \eqref{eq:upd1}; \label{l:l1}
		\State $\mv{\Sigma}^{-1}\leftarrow\mv{\Sigma}^{-1}-\eta\frac{\mv{\Sigma}^{-1}\mv{s}_{n}\mv{s}^H_{n}\mv{\Sigma}^{-1}}{1+\eta\mv{s}^H_{n}\mv{\Sigma}^{-1}\mv{s}_{n}}$;\label{l:l2}
		\EndFor\label{l:lN}
		\State Compute $\mv{V}\left(\mv{\gamma}\right)$ according to \eqref{V2};
		\EndWhile
		\State Output: $\mv{\gamma}$.
	\end{algorithmic}
\end{algorithm}

\subsection{Review of Coordinate Descent (CD) Algorithm in \cite{haghighatshoar2018improved}} 

The activity detection problem in \eqref{eq:P6} has been popularly solved by the CD algorithm. In this subsection, we first review an existing CD algorithm in \cite{haghighatshoar2018improved} since our proposed algorithm by exploiting $\hat{K}$ is building on that CD algorithm.

In the CD algorithm, at the $i$-th iteration, the CD algorithm updates each coordinate as follows:
\begin{align}
	\gamma_n\leftarrow\gamma_n+\eta, ~ n=1, 2,\dots, N,\label{eq:upd1}
\end{align}
where 
\begin{align}
	\small{\!\eta\!=\!\min\left\{\!\max\! \left\{\!\frac{\mv{s}^H_{n}\mv{\Sigma}^{-1}\hat{\mv{\Sigma}}\mv{\Sigma}^{-1}\mv{s}_{n}-\mv{s}^H_{n}\mv{\Sigma}^{-1}\mv{s}_{n}}{\left(\mv{s}^H_{n}\mv{\Sigma}^{-1}\mv{s}_{n}\right)^2},-\gamma_n\right \}\!,1\!-\!\gamma_n\right\}.}\label{eta}
\end{align}
The algorithm is terminated when 
\begin{align}
	\|\mv{V}\left(\mv{\gamma}\right)\|_{\infty}\le\varepsilon,\label{eq:ter}
\end{align}
where $\mv{V}\left(\mv{\gamma}\right)=[V(\gamma_1),V(\gamma_2),\dots,V(\gamma_N)]^T$, $\|\mv{V}\left(\mv{\gamma}\right)\|_{\infty}$ denotes the infinity norm of $\mv{V}\left(\mv{\gamma}\right)$, and $\varepsilon$ denotes the error tolerance.  
The CD algorithm is summarized in Algorithm \ref{alg:RCD}.

\subsection{Proposed K-Coordinate Descent Algorithm} \label{sec:kCD} 
A potential problem in the CD algorithm is its high complexity of updating all coordinates at each iteration (i.e.,  Lines \ref{l:l0}--\ref{l:lN} in Algorithm \ref{alg:RCD}). Specifically, the complexity of updating one coordinate is in the order of $\mathcal{O}(L_{\rm II}^2)$. Therefore, the total per-iteration complexity of updating $N$ coordinates is in the order of $\mathcal{O}(NL_{\rm II}^2)$. In particular, when $N$ is large, the CD algorithm will suffer from a huge complexity.

In the following, we exploit the information $\hat{K}$ obtained in Phase I to reduce the per-iteration computational complexity in CD. Specifically, through the estimation of $K$ in Phase I, we have high confidence in believing that there are $\hat{K}$ active devices, and updating the coordinates in $\mv{\gamma}$ corresponding to these $\hat{K}$ active devices can mostly reduce the objective function value of problem  \eqref{eq:b}. In this case, we only need to update at most $\hat{K}$ coordinates in the CD algorithm at each iteration. As a result, the per-iteration complexity of our proposed CD algorithm  is in the order of $\mathcal{O}(\hat{K}L_{\rm II}^2)$, which is significantly smaller than that in \cite{haghighatshoar2018improved} when $\hat{K}$ is far smaller than $N$ (a very common scenario in mMTC). Below we show the details of our proposed CD algorithm by judiciously exploiting the information $\hat{K}$ obtained in Phase I.

\textbf{Selecting only $\hat{K}$ coordinates to update:} We propose a criterion based on the first-order optimality metric in \eqref{V2} to update at most $\hat{K}$ coordinates among the total $N$ coordinates at each iteration.  Specifically, as introduced before, a larger $V(\gamma_n)$ in \eqref{V2} indicates a larger violation of the first-order optimality condition, $n=1, 2, \dots, N$. In this case, at the $i$-th iteration of Algorithm \ref{alg:RCD}, we first sort the components in $\mv{V}(\mv{\gamma})$ in the descending order as follows:
\begin{align}
	V(\gamma_{i_1})\ge V(\gamma_{i_2})\ge\dots\ge V(\gamma_{i_N}); \label{eq:s}
\end{align}
Then, we define an importance index set $\mathcal{A}^{(i)}$ that contains the indices of the $\hat{K}$ largest $V(\gamma_n)$ as follows:
\begin{align}
	\mathcal{A}^{(i)}=\{i_1,i_2,\dots,i_{\hat{K}}\}; \label{eq:A}
\end{align}
Finally, we only update the coordinates with $n\in\mathcal{A}^{(i)}$. Thus, the per-iteration complexity of Algorithm \ref{alg:RCD} can be reduced significantly. 

\textbf{Reducing the complexity in computing $\mv{V}(\mv{\gamma})$:} In order to get the importance set $\mathcal{A}^{(i)}$ in \eqref{eq:A}, we need to compute $V(\gamma_n)$ in \eqref{V2} and $\left[\nabla f(\mv{\gamma})\right]_n$ in \eqref{eq:grad}, $n=1,2,\dots, N$. The computation of $\left[\nabla f(\mv{\gamma})\right]_n$ will cause additional complexity,  which is actually in the same order of the complexity when updating a coordinate in CD.  The motivation here  is to spend less computational cost while achieving the coordinate selection as in  $\mathcal{A}^{(i)}$. In this way, we can further reduce the per-iteration computational complexity of the CD  algorithm.

Observe that, if $V(\gamma_n)$ is below a threshold for several iterations, the coordinate $\gamma_n$ has likely converged. In this case, there is no need to update $\gamma_n$ in  subsequent iterations. To leverage the above observation, we define a quantity $b_n$ for coordinate $\gamma_n$, which indicates how many times the violation $V\left(\gamma_n^{(i)}\right)$ of the coordinate is below a predefined threshold $\alpha$, i.e., 
\begin{align}
	b_n=\left\{\begin{aligned}
			&b_n+1,\   ~{\rm if}~V\left(\gamma_n^{(i)}\right)<\alpha, \\
			&b_n,\   ~~~~~~{\rm otherwise}, 
		\end{aligned}\right.~n=1,2,\dots, N,\label{Count}
\end{align}
where $\gamma_n^{(i)}$ denotes the value of  $\gamma_n$ at the $i$-th iteration. The initial value of $b_n$ is set to be zero. Then, the candidate coordinate set to be updated at the $i$-th iteration is defined as
\begin{align}
	\mathcal{C}^{(i)}=\{n \mid b_n\le D\},\label{SetE}
\end{align}
where $D$ is an integer that controls how many times that $V(\gamma_n)$ falls below the threshold $\alpha$ is allowed.  If $n\notin\mathcal{C}^{(i)}$, in the subsequent iterations, we no longer compute the gradient $\left[\nabla f(\mv{\gamma})\right]_n$ and 	$V(\gamma_n)$.

\textbf{Refining set $\mathcal{A}^{(i)}$ in \eqref{eq:A}:} Through the design of coordinate set $\mathcal{C}^{(i)}$ in \eqref{SetE},  the coordinates $n\notin\mathcal{C}^{(i)}$ will not be updated anymore in the rest of the iterations. Accordingly, we only need to compute and sort $V(\gamma_n)$ for $n\in\mathcal{C}^{(i)}$, i.e.,
\begin{align}
	V(\gamma_{i_1})\ge V(\gamma_{i_2})\ge\dots\ge V(\gamma_{i_{E}}),
\end{align}
where 
\begin{align}
	E=\min\{|\mathcal{C}^{(i)}|,\hat{K}\},\label{eq:E}
\end{align}
and $|\mathcal{C}^{(i)}|$ denote the cardinality of $\mathcal{C}^{(i)}$. Then,  $\mathcal{A}^{(i)}$ in \eqref{eq:A} is refined as follows:
\begin{align}
	\mathcal{A}^{(i)}=\{i_1,i_2,\dots,i_{E}\}. \label{eq:A2}
\end{align}

Based on the above discussion, we can see that only at most $\hat{K}$ coordinates are updated at each iteration,  we name this algorithm as K-coordinate descent (K-CD). The K-CD algorithm is summarized in Algorithm \ref{alg:FCD}.

\begin{algorithm}[t]
	\caption{Proposed K-CD Algorithm}\label{alg:FCD}
	\begin{algorithmic}[1] 
		\State Input: $\mv{\gamma}=\mv{0}$,  $\|\mv{V}\left(\mv{\gamma}\right)\|_{\infty}$, $\mv{\Sigma}=\sigma^2 \mv{I}$, $\varepsilon$, $\alpha$, $i=0$, $D$.
		\While {$ \lVert \mv{V}\left(\mv{\gamma}\right) \rVert_{\infty}>\varepsilon$}
		\State $i\leftarrow i+1$;
		\State Update $\mathcal{C}^{(i)}$ according to \eqref{SetE};
		\State Update $\mathcal{A}^{(i)}$  according to \eqref{eq:A2};
		\State Run Lines \ref{l:l1} and \ref{l:l2} in Algorithm \ref{alg:RCD} for $n\in \mathcal{A}^{(i)}$;
		\State Compute $V(\gamma_n)$ for $n\in \mathcal{C}^{(i)}$;
		\EndWhile 
		\State Output: $\mv{\gamma}$. 
	\end{algorithmic}
\end{algorithm} 

\subsection{A Related Work in \cite{wang2023cov}}
Prior to our work, the idea of reducing the per-iteration computational complexity by updating a selected subset of coordinates at each iteration of CD was developed in \cite{wang2023cov}, where the algorithm is referred to as Active Set CD. In Active Set CD, the importance index set is 
\begin{align}
	\mathcal{A}^{(i)}=\big{\{}n \ | \ V\big{(}\gamma_n^{(i)}\big{)} \geq \omega\big{\}},\label{eq:Ai}
\end{align} 
where $\omega$ is a pre-selected constant, and the coordinate set is  $\mathcal{C}^{(i)}=\{1, 2,\dots, N\}$.

There are two key differences between our proposed K-CD algorithm and the Active Set CD algorithm in \cite{wang2023cov}. First, the criterion for selecting which coordinate to update in Active Set CD is through the preset parameter $\omega$. In sharp contrast, our K-CD algorithm selects coordinates based on an estimation of $K$. Numerical results in Section \ref{Results} will demonstrate that K-CD significantly reduces the number of coordinate updates compared with Active Set CD. Second, unlike Active Set CD which needs to compute the full gradient at each iteration, K-CD only needs to compute a partial gradient at each iteration, whose cardinality is generally significantly less than the problem's dimension $N$. This reduction plays a significant role in reducing the computational complexity, as the computation of the gradient itself is a major source of complexity.

\subsection{Complexity Analysis} \label{sec:com}
We compare the per-iteration complexity of the proposed K-CD algorithm with that of  CD and Active Set CD algorithms in terms of floating point operations (FLOPs). The per-iteration complexity of the proposed K-CD algorithm is
\begin{align}
 5L_{\rm II}^2E+4L_{\rm II}^2\vert\mathcal{C}^{(i)}\vert,
\end{align}
where  $5L_{\rm II}^2$ is the complexity of one coordinate update, and $4L_{\rm II}^2$ is the complexity  of one gradient component computation (i.e., the complexity of computing $\left[\nabla f(\mv{\gamma})\right]_n$ in \eqref{grad equa2}), $E$ is given  in \eqref{eq:E} and $\mathcal{C}^{(i)}$ is given in \eqref{SetE};  the per-iteration complexity of the
CD algorithm is $5L_{\rm II}^2N$; the per-iteration complexity of the Active Set CD algorithm is $5L_{\rm II}^2\vert\mathcal{A}^{(i)}\vert+4L_{\rm II}^2N$, where $\mathcal{A}^{(i)}$ is given in \eqref{eq:Ai}.


\begin{figure*}[t]
	\begin{minipage}[t]{0.3\textwidth}
		\centering
		\includegraphics[width=\linewidth]{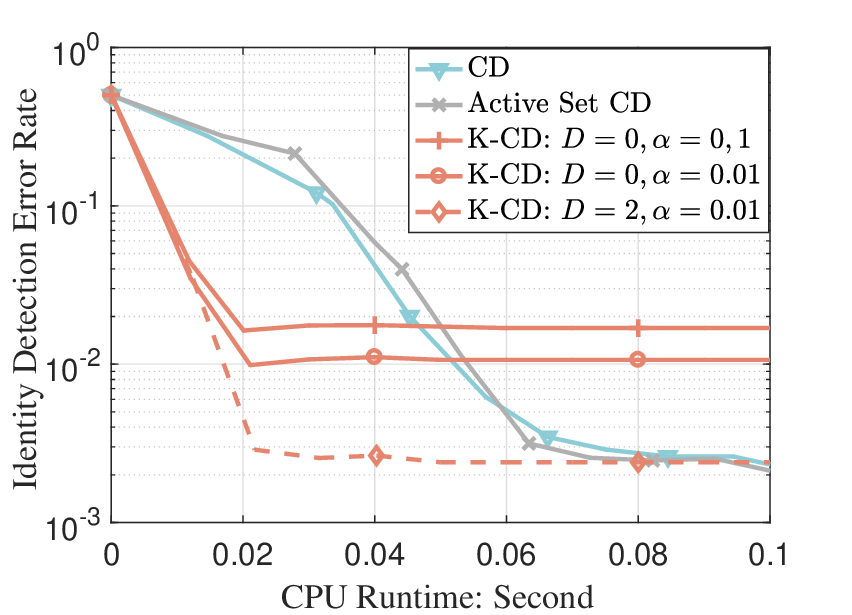}
		\caption{Activity detection performance when $K=100$, $L_{\rm I}=4$, $L_{\rm II}=100$, and $M=15$.}
		\label{Per}
	\end{minipage}
	\hspace{6mm}
	\begin{minipage}[t]{0.3\textwidth}
		\centering
		\includegraphics[width=\linewidth]{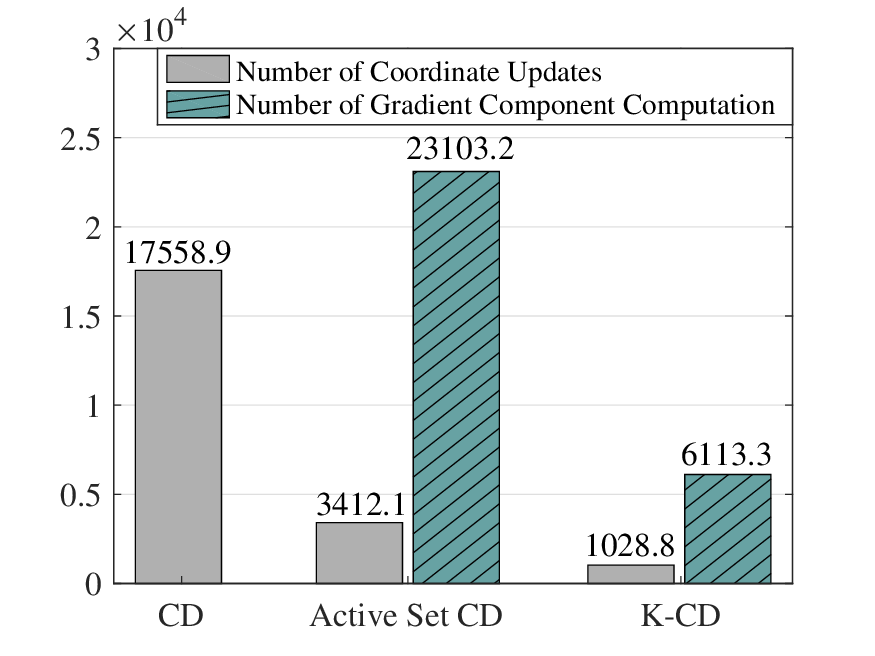}
		\caption{Number of total coordinate updates and gradient component computation when algorithms meet the  termination condition.}
		\label{Com}
	\end{minipage}
	\hspace{6mm}
	\begin{minipage}[t]{0.3\textwidth}
		\centering
		\includegraphics[width=\linewidth]{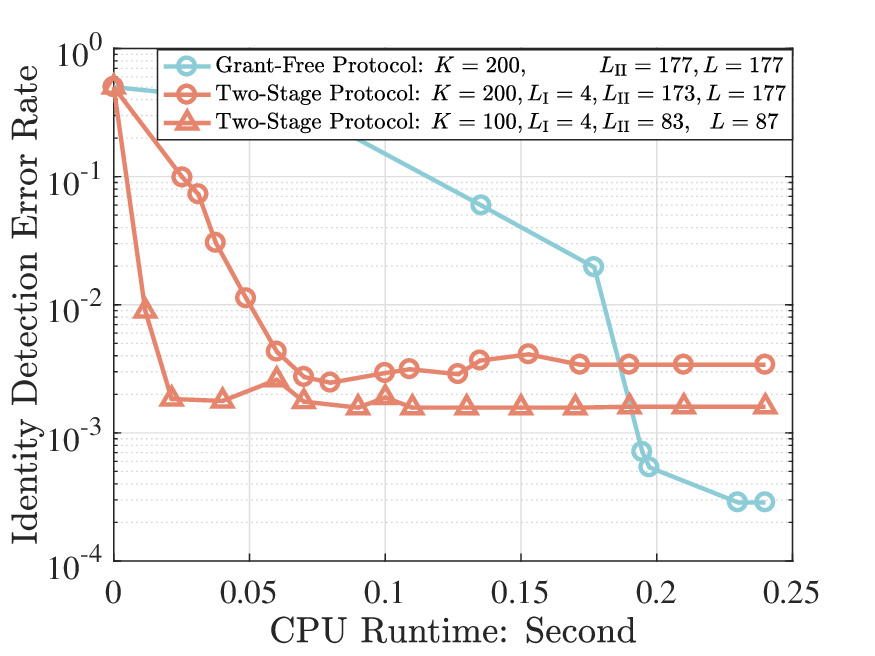}
		\caption{Performance of the proposed two-stage protocol when $M=32$.}
		\label{2stage}
	\end{minipage}
\end{figure*}

\section{Numerical Results}\label{Results}
In this section, we show the performance of the identity detector for active devices, and the overall performance of the proposed two-stage communication protocol in terms of the identity detection error rate, which is the error rate when the missed detection probability (MDP) equals the false alarm probability (FAP). The MDP is the probability that an active device is detected to be inactive; and the FAP is the probability that an inactive device is detected to be active. In the simulations, we set $N=1000$ and $\varepsilon=10^{-3}$. The channel setup is the same as that in Section \ref{sec:D}. 
\subsection{Performance of K-CD Algorithm}
We have two benchmarks to evaluate the identity detection performance of the proposed K-CD algorithm, which are CD algorithm\cite{haghighatshoar2018improved} and Active Set CD algorithm\cite{wang2023cov}.

In Figs. \ref{Per} and \ref{Com}, we show the performance of the proposed K-CD identity detector designed in Phase II when $K=100$ and $M=15$. In the simulations, we first allocate $L_{\rm I}=4$ symbols to estimate $K$ by applying the designed estimator in Phase I. The output $\hat{K}$ is then fed to the detector in Phase II, where we allocate $L_{\rm II}=100$ symbols to detect the identities of active devices. In addition, to have a fair comparison, we do not count the CPU runtime when computing $V(\mv{\gamma})$ in the CD algorithm while we do count the CPU runtime when computing $V(\mv{\gamma})$ in the K-CD and Active Set CD algorithms. Fig. \ref{Com} shows the average total number of coordinate updates and gradient component computation when the algorithms meet the termination condition in \eqref{eq:ter}. 

First, as shown in Fig. \ref{Per}, with a sufficient runtime, K-CD with $D=2$ and $\alpha=0.01$, alongside CD and Active Set CD all achieve the same detection performance. This shows that although K-CD reduces the computational complexity, K-CD does not compromise the detection performance. Second, compared with the CD algorithm, our K-CD algorithm notably reduces the convergence time by a factor of three. The reason behind is shown in Fig. \ref{Com}. Specifically, K-CD reduces the total number of updated coordinates significantly compared with CD. Although  K-CD needs to compute some partial gradients additionally, the overall complexity is reduced in the end. Third, the computational complexity of the K-CD algorithm is significantly less than that of Active Set CD. This is due to the fact that, as shown in Fig. \ref{Com}, K-CD can reduce the number of coordinate updates and gradient component computation significantly at the same time, while Active Set CD only reduces the number of coordinate updates. Last, if we fix the runtime to be 0.02s, the detection performance of K-CD outperforms CD and Active Set CD by nearly two orders of magnitude.  

\begin{table}[t]
	\centering
	\setlength{\tabcolsep}{10pt}
	\caption{Relationship between $K$ and $L_{\rm II}$ when the threshold is $10^{-2}$, $M=32$, and $N=1000$}
	\label{tb:map}
	\begin{tabular}{|c|c|}
		\hline
		$K$ &  $L_{\rm II}$\\ \hline
		10 & 15 \\ \hline
		20 & 25 \\ \hline
		30 & 35 \\ \hline
		40 & 45 \\ \hline
		50 & 55 \\ \hline
		60 & 65 \\ \hline
		70 & 75 \\ \hline
		80 & 76 \\ \hline
		90 & 77 \\ \hline
		100 & 78 \\ \hline
	\end{tabular}
	\begin{tabular}{|c|c|}
		\hline
		$K$ &  $L_{\rm {\uppercase\expandafter{\romannumeral 2}}}$ \\ \hline
		110 & 80 \\ \hline
		120 & 90 \\ \hline
		130 & 100 \\ \hline
		140 & 110 \\ \hline
		150 & 120 \\ \hline
		160 & 130 \\ \hline
		170 & 140 \\ \hline
		180 & 150 \\ \hline
		190 & 160 \\ \hline
		200 & 170 \\ \hline
	\end{tabular}
	\begin{tabular}{|c|c|}
		\hline
		$K$ &  $L_{\rm {\uppercase\expandafter{\romannumeral 2}}}$ \\ \hline
		210 & 180 \\ \hline
		220 & 190 \\ \hline
		230 & 200 \\ \hline
		240 & 210 \\ \hline
		250 & 220 \\ \hline
		260 & 230 \\ \hline
		270 & 240 \\ \hline
		280 & 250 \\ \hline
		290 & 260 \\ \hline
		300 & 270 \\ \hline
	\end{tabular}
\end{table}
\subsection{Performance of Two-Stage Communication Protocol}
We compare the proposed two-stage communication protocol with the existing grant-free protocol, which applies the CD algorithm (Algorithm \ref{alg:RCD}) for activity detection.

In the two-stage protocol, we apply Table \ref{tb:map} to determine $L_{\rm II}$ based on $\hat{K}$. The table is constructed according to a  predefined performance threshold, i.e., the maximum identity detection error rate of the two-stage protocol, which is set as $10^{-2}$. Even though $\hat{K}$ is only an estimate value of $K$, we proceed as if it were accurate when referring Table \ref{tb:map}. The table covers a range of $K$ from 10 to 300 with increments of 10. If $\hat{K}$ does not exactly match a table boundary but falls within an interval, $L_{\rm II}$ is assigned based on the higher boundary of that interval. For example, when  $\hat{K}=10$, $L_{\rm II}=15$, and when $10<\hat{K}<20$, $L_{\rm II}=25$. In addition, since the two-stage protocol allocates  $L_{\rm II}$ dynamically, in Fig. \ref{2stage}, we still use $L_{\rm II}$ to denote the average allocated preamble length in Phase II.

First, Fig. \ref{2stage} shows that with a sufficient runtime the proposed two-stage protocol achieves a stable identity detection performance as $K$ increases from 100 to 200, keeping the detection error rate below the predefined threshold $10^{-2}$. This is in sharp contrast to the grant-free protocol, where increasing $K$ from 100 to 200 significantly degrades the detection performance. Second, for  $K=200$, when the grant-free protocol is allocated $L = L_{\rm II} = 177$ preamble symbols which is equal to the total average preamble symbols $L=177$ in the two-stage protocol, Fig. \ref{2stage}  shows that the performance of the proposed two-stage protocol is much better than that of the grant-free protocol when the runtime is smaller than 0.15s. The performance gain comes from the exploitation of $\hat{K}$ in Phase II to reduce  complexity.

\bibliographystyle{IEEEtran}
\bibliography{CIC}

\end{document}